\documentstyle[preprint,aps]{revtex}
\begin{document}
\draft
\begin{title}
{
Logarithmic divergence of normal state resistivity of bipolaronic
high-$T_{c}$ cuprates.
}
\end{title}
\author
{A.S. Alexandrov}
\address{
Department of Physics, Loughborough University, Loughborough
LE11 3TU, U.K.
}
\maketitle
\begin{abstract}
The resonance Wigner scattering  of charged bosons (small bipolarons) in a
random potential
leads to
  logarithmically divergent low-temperature resistivity
 as observed in several
cuprates.
Unusual temperature dependence of resistivity of $La_{2-x}Sr_{x}CuO_{4}$
 as well as of the Hall effect
 is  quantitatively described  in a wide temperature range providing
an evidence for $2e$ charged Bose-liquid in high-$T_{c}$
cuprates.
\end{abstract}

\narrowtext

There is no need to abandon the Boltzmann kinetics to explain the
linear in-plane resistivity and temperature dependent Hall effect above
$T_{c}$
in cuprates if the bipolaron theory \cite{alemot} is applied.
A fraction of the bipolarons is localised by
disorder, so that the number of delocalised carriers is
proportional to $T$ while the boson-boson inelastic scattering rate is
proportional to $T^{2}$\cite{ale}. This allows us to explain that both the
in-plane resistivity and the Hall density are proportional to $T$.
 Recently we have extended the theory towards  $c$-axis transport
\cite{ale2}. The
semiconductor-like
temperature dependence of the $c$-axis resistivity  has been
understood within the
Boltzmann kinetics   by taking into account the contribution of
thermally excited polarons to the $c$-axis transport.

In this letter  the  transport relaxation time of
bipolarons (or any charged bosons on a lattice) is derived at low
temperatures
where the scattering
 is dominated by the elastic boson-impurity scattering andsingle polarons
are frozen out. The temperature dependence of
the resistivity appears to be in a remarkable agreement with the
experimental dependence measured recently by Ando $et$ $al$ \cite{ando}
in $La_{2-x}Sr_{x}CuO_{4}$ by suppressing $T_{c}$ down to a
millikelvin scale with a pulsed magnetic field.

Small hole polarons are
    paired  $above$ $T_{c}$
    in the doped Mott-Hubbard insulators  if the electron-phonon
    interaction is relatively strong ($\lambda >0.5$)\cite{alemot}.
   Therefore, the low-energy states of cuprates are thought to be a
   mixture of the intersite in-plane singlet  pairs (small bipolarons) and
   thermally excited polarons.
   Above $T_{c}$,
   which is the condensation temperature of the charged Bose-gas,
  these carriers are nondegenerate.  Intersite singlets
   tunnel along the planes with an effective mass $m^{**}$ of the order
   of a single-polaron mass \cite{aleA}.
   Their $c$-axis tunneling is
   Josephson-like  involving the simultaneous hopping of two holes.
   Therefore the singlet $c$ -axis mass is strongly enhanced,
   $m^{**}_{c}\gg m^{**}$,  which leads to a
   large transport
   anisotropy  at low temperatures when  polarons are
   frozen out. At  temperatures below the $c$-axis
   bipolaron bandwidth  we expect  a three-dimensional anisotropic energy
   spectrum and a three dimensional  scattering of bipolarons
   dominated by the lattice defects and
   impurities. The number of extended bosons $n_{b}(T)$ above
   the mobility edge is determined in the `single-well---single-particle'
approximation
   as \cite{ale}
   \begin{equation}
    n_{b}(T)={x\over{2}}-n_{L}(T),
\end{equation}
    where $x/2$ is the total number of pairs, and
    $n_{L}(T)\simeq n_{L}-N_{L}(0)k_{B}T$ is the number of bosons
    localized by the random potential with $N_{L}(0)$ the density of
    localized states near the mobility edge. The Hall
    coefficient $R_{H}$ measures the inverse carrier density , so that
    \begin{equation}
    {R_{H}\over{R_{H0}}}={1\over{1+T/T_{L}}},
    \end{equation}
    where $T_{L}=(x-2n_{L})/2k_{B}N_{L}(0) $. This simple expression fits
     well to the Hall coefficient temperature dependence of
    $La_{2-x}Sr_{x}CuO_{4}$ at optimum doping
    ($x=0.15$)\cite{bat} as
    shown in Fig.1 with $T_{L}=234K$ and  the constant $R_{H0}=2\times
10^{-3} cm^{3}/C$.
    If the total number of carriers $x/2$ is above the total number of the
    potential wells $n_{L}$, which is assumed here, the carrier
    density is practically temperature independent at low
    temperatures \cite{ando2}. On the other hand if $x < 2n_{L}$ the
    re-entry effect into the normal state  appears with the
    temperature lowering below $T_{c}$\cite{ale3}.

    The normal  state of the
    bipolaronic superconductor is reminiscent to that of a nondegenerate
    semiconductor. The characteristic kinetic energy of carriers
    appears to be of the order of the temperature rather than of
     the Fermi energy of usual metals. The most
    effective scattering at low temperatures is then caused by the
attractive
    shallow
    potential wells which for slow particles is described by
    the familiar Wigner
    resonance cross-section \cite{lan}
    \begin{equation}
 \sigma(E)={2\pi \hbar^{2}\over{m}}{1\over{E+|\epsilon|}}.
    \end{equation}
 Here
 \begin{equation}
\epsilon
=-{\pi^{2}\over{16}}U_{min}\left({U\over{U_{min}}}-1\right)^{2}
\end{equation}
is the energy of a shallow virtual ($U < U_{min}$) or real ($U > U_{min}$)
localised level, $U$  the well depth and
$U_{min}=\pi^{2}\hbar^{2}/8ma^{2}$ with  the well size $a$. In
the non-crossing approximation, which  describes  the essential
physics, the transport relaxation rate is the sum of the scattering
cross-sections from different potential wells within the unit volume
multiplied by the velocity $v=\sqrt{2E/m}$. There is a wide
distribution of potential wells with respect to both $U$ and $a$ in
real cuprates.
Therefore, one has to integrate the Wigner cross-section, Eq.(3),
over $U$ and over $a$. By doing the integration over $U$ we take into
account only shallow wells with $U<U_{min}$ because the deeper wells
are occupied by  localised carriers and cannot yield a resonant
scattering. The result is
\begin{equation}
\langle \sigma(E)\rangle \equiv \gamma^{-1}\int_{0}^{U_{min}}\sigma(E)dU
={4\pi \hbar^{3}\over{m\gamma a
\sqrt{2mE}}}tan^{-1}\left({\pi^{2}\hbar\over{8a\sqrt{2mE}}}\right),
\end{equation}
where  the width of the $U$-distribution $\gamma$ is supposed to be
large compared with $U_{min}$. By doing the integration over size $a$
one has to take into account the fact that the Wigner formula, Eq.(3) is
applied
to slow particles with $a\leq \hbar/\sqrt {2mE}$. However,
because the $U$-averaged cross-section, Eq.(5) behaves like $1/a^{2}$ at
large $a$ one can  extend the integration over $a$ up to infinity to
obtain the inverse mean free path as
\begin{equation}
l^{-1}(E) = {n_{L}\over{A}}\int_{a_{min}
}^{\infty}\langle\sigma(E)\rangle da\simeq
{\pi^{2}\hbar^{3}N_{L}(0)\over{m A\sqrt{2mE}}}\ln{E_{0}\over{E}}
\end{equation}
for $E\ll E_{0}$. Here $A$ is the width of the size distribution of the
random
potential, $N_{L}(0)=n_{L}/\gamma $ \cite{ale},
$E_{0}=\pi^{4}\hbar^{2}/128 ma_{min}^{2}$  and $a_{min}$
is the minimum size. We expect a very large
value of $A$ of order of a few tens $\AA$ due to the twin
boundaries and impurity clusters in real cuprates which are not
screened. On the other hand,  single impurities are screened. A simple
estimate of the screening radius by the use of the classical expression,
 $r_{D}=\sqrt{k_{B}T\epsilon /16 \pi
n_{B}e^{2}}$ yields a value of $a_{min}$ of the order or even less than
the interatomic spacing ($\sim 1.9\AA$), which corresponds to a quite
large
$E_{0}/k_{B}=1500K$ if $m=10 m_{e}$. As a result, in a wide temperature
range we arrive at the
logarithmic transport relaxation rate as
\begin{equation}
{1\over{\tau}}\equiv v l^{-1}(E)={1\over{\tau_{0}}}\ln{E_{0}\over{E}},
\end{equation}
where $\tau_{0}=m^{2}A/\pi^{2}\hbar^{3}N_{L}(0)$ is a constant.
The above formula is applied to the isotropic s-scattering
of slow
particles with an isotropic $3D$
energy spectrum described by the effective mass $m$. However,
assuming that the attractive field scales as
$U({\bf r})=U(x^{2}+y^{2}+z^{2}m^{**}_{c}/m^{**})$ one obtains  the same
result  for
the anisotropic spectrum as well,
if  $m=m^{**}$. Because the Wigner formula, Eq.(3) is somewhat
more general than the assumption made in its derivation \cite{lan} we
 expect that the obtained logarithmic dependence, Eq.(7) is not changed if
the attraction potential $U({\bf r})$ is modified.

The low temperature resistivity is now derived by the use of
Boltzmann theory as
\begin{equation}
\rho(T)= \rho_{0}\ln { \omega_{0}\over{T}},
\end{equation}
where $\rho_{0}^{-1}=2(x-2n_{L})e^{2}\tau_{0}/m$ and
$\omega_{0}=E_{0}/k_{B}$.
 At high temperatures the inelastic scattering
of extended bosons by localised bosons becomes important as discussed
in ref.\cite{ale} so that
\begin{equation}
{1\over{\tau}}=\alpha T^{2}
\end{equation}
with the constant, $\alpha$,  proportional to the density of states at the
mobility edge,
$N_{L}(0)$ squared. Combining both elastic, Eq.(7), and inelastic, Eq.(9),
 scattering, and taking into account the temperature dependence of
the extended boson density $n_{B}(T)$, Eq.(1), we  arrive at
\begin{equation}
{\rho(T)\over{\rho_{0}}}={\ln( \omega_{0}/T) +
(T/T_{B})^{2}\over{1+T/T_{L}}},
\end{equation}
with the constant $T_{B}=1/\sqrt{\alpha \tau_{0}}$.
The solid line in Figs.2,3
 is a fit to the
experimental data  with $\rho_{0}=7.2 \times 10^{-5} \Omega
cm$,  $\omega_{0}=1900K$\cite{ando} and $T_{B}=62K$ which appears to be
remarkably good. The value of  $\omega_{0}$
agrees
well with the  estimate above. In the underdoped samples the effective
mass of bipolarons as well as the screening radius significantly
increases  \cite{alemot}, so $\omega_{0}$ might be less about one order of
magnitude.
On the other hand,  the logarithmic dependence derived above is well
verified  in the temperature range $\hbar^{2}/2mA^{2}< k_{B}T<
\hbar^{2}/2ma_{min}^{2}$. With increasing doping the screening becomes
more efficient and   potential wells of a large size $A$ are less
probable. Hence, the temperature interval for the logarithmic dependence
is less, and its disappearance in  overdoped samples is expected.
 Because the bipolaron energy spectrum is three dimensional at low
 temperatures, there is
 no temperature dependence of the anisotropy $\rho_{c}/\rho_{ab}$ at low
$T$ as
observed \cite{ando}.

A simple estimate of the polaronic level shift in $LSCO$
shows that the electron-phonon coupling is more than sufficient to bind
two polarons into a small bipolaron. For the Fr\"ohlich
interaction  the polaron level shift   is given by \cite{aleA}
\begin{equation}
E_{p}\simeq
{q_{D}e^{2}\over{\pi}}(\epsilon^{-1}-\epsilon^{-1}_{0}),
\end{equation}
where $q_{d}=(6\pi^{2}/\Omega)^{1/3}$ is the Debye momentum, $\Omega$ the
unit cell volume. With the
static and high frequency dielectric constants
$\epsilon_{0}>>\epsilon \simeq 5$ and $q_{D}\simeq 0.7\AA^{-1}$
one obtains $E_{p}\simeq 0.64eV$. As a
result, there is a  large attraction between two polarons  of the order of
$2E_{p}\simeq 1eV$. At the same time  the band mass enhancement
is less than one order of magnitude because of the phonon dispersion as
shown in ref.\cite{aleA}.  This is in contrast with
  some assessments of the electron-phonon interaction and the bipolaronic
mechanism of high-$T_{c}$ superconductivity based on the incorrect
estimate of the coupling constant and of the effective mass.
 The hole-hole coupling via phonons, Eq.(11),
is at least
four times stronger than the magnetic coupling,  determined by
the exchange  $J\simeq 0.15 eV$, which makes
any model of
high-$T_{c}$ without phonons highly unrealistic (see also ref.\cite{mul}).
 A large isotope effect on both $T_{c}$
  \cite{fra} and on the N\'eel temperature $T_{N}$ \cite{mor} in $LSCO$
confirms
  this conclusion.  The pair-distribution analysis of neutron
  scattering \cite{sen}  suggests that the `spin-gap' is consistent
  with the formation of a bipolaronic local singlet state as we predicted
  well before high-$T_{c}$ cuprates were discovered \cite{aleran}. We
  believe  that the
  normal state `pseudogap' observed with magnetic, kinetic, thermodynamic
  and $ARPES$ measurements  has the same
  origin \cite{aleA,alemot}, and a
  so-called ` pairing without phase coherence' \cite{cam} is none other
  than the bipolaron formation.

 In conclusion, the bipolaron kinetics  remarkably well describes
 the unusual logarithmic divergence of the low-temperature
 resistivity as well as the  high-temperature resistivity and
 the  Hall effect observed in several high-$T_{c}$ cuprates,
 in particular, in
 the reference system $La_{2-x}Sr_{x}CuO_{4}$\cite{ando}. The logarithmic
 divergence is caused by the resonance Wigner scattering of slow
 nondegenerate carriers (bipolarons) scattered by the
 attractive random potential in the $3D$ doped insulators.

  I highly appreciate the enlightening discussions with Sir Nevill
  Mott, Y. Ando, A. Bratkovsky,
  P.P.  Edwards, J. Cooper, R. Haydock, N.
  Hussey, V. Kabanov, D. Khmelnitskii, W. Liang,  J. Loram,  and K.R.A.
Ziebeck.

\figure{The Hall effect in $La_{2-x}Sr_{x}CuO_{4}$  (triangles\cite{bat})
described by the theory\cite{ale} (solid line) for $x=0.15$.}

\figure {The `ab' resistivity of $La_{2-x}Sr_{x}CuO_{4}$ with
$x=0.13$  (diamonds\cite{ando}) described by Eq.(7) in a wide temperature
range (solid line).}
\figure {The same as in Fig.2 at  low temperatures.}
 \end{document}